\documentclass[fleqn,usenatbib]{mnras}

\usepackage{float}
\usepackage{xcolor}

\newcommand{\Ayr}{A_{\mathrm{yr}}}

\newcommand{\logno}{\log_{10}{\dot n}_0}
\newcommand{\no}{{\dot n}_0}
\newcommand{\alphaM}{\alpha_{\mathcal{M}}}
\newcommand{\betaz}{\beta_{z}}
\newcommand{\Mchirp}{\mathcal{M}}
\newcommand{\Mstar}{\mathcal{M}_*}
\newcommand{\Msol}{M_{\odot}}

\newcommand{\lognMLowPost}{-10.9}

\newcommand{\lognMHighPost}{0.3}

\definecolor{barcelonafcgold}{HTML}{edbb00}

\DeclareRobustCommand{\VAN}[3]{#2}
\let\VANthebibliography\thebibliography
\def\thebibliography{\DeclareRobustCommand{\VAN}[3]{##3}\VANthebibliography}

\usepackage{graphicx}	
\usepackage{amsmath}	
\usepackage{amssymb}	
\usepackage{tabularx}

\usepackage{comment}

\title[MBHBs and the NANOGrav 12.5 years results]{Massive black hole binary systems and the NANOGrav 12.5 year results}

\author[Middleton et al.]{
H.~Middleton,$^{1,2}$
A.~Sesana,$^{3,4}$
S.~Chen,$^{5,6}$
A.~Vecchio,$^{7}$
W.~Del Pozzo$^{8}$
and P.A.~Rosado$^{9}$
\\
\\
$^{1}$School of Physics, University of Melbourne, Parkville, Victoria 3010, Australia\\
$^{2}$OzGrav-Melbourne, Australian Research Council Centre of Excellence for Gravitational-wave Discovery, University of Melbourne, \\ Parkville, Victoria 3010, Australia\\
$^{3}$Department of Physics ``G. Occhialini'', University of Milano - Bicocca, Piazza della Scienza 3, 20126 Milano, Italy\\
$^{4}$INFN Sezione di Milano - Bicocca, Piazza della Scienza 3, 20126 Milano, Italy\\
$^{5}$Laboratoire de Physique et Chimie de l'Environnement et de l'Espace, Universit\'{e} d'Orl\'{e}ans, CNRS, 45071 Orl\'{e}ans, France \\
$^{6}$Station de Radioastronomie de Nan\c{c}ay, Observatoire de Paris, PSL University, CNRS, 18330 Nan\c{c}ay, France \\
$^{7}$School of Physics and Astronomy \& Institute for Gravitational Wave Astronomy, University of Birmingham, Birmingham, B15 2TT, UK\\
$^{8}$Dipartimento di Fisica ``Enrico Fermi'', Università di Pisa, Pisa I-56127, Italy and INFN sezione di Pisa, Pisa I-56127, Italy\\
$^{9}$Holaluz-Clidom S.A., Passeig de Joan de Borbó 99-101, 4a Planta, 08039 Barcelona, Spain
}

\date{Accepted XXX. Received YYY; in original form ZZZ}

\pubyear{2020}

\begin{document}
\label{firstpage}
\pagerange{\pageref{firstpage}--\pageref{lastpage}}
\maketitle

\begin{abstract}
The North American Nanohertz Observatory for Gravitational Waves (NANOGrav) has recently reported evidence for the presence of a common stochastic signal across their array of pulsars. The origin of this signal is still unclear. One of the possibilities is that it is due to a stochastic gravitational wave background (SGWB) in the $\sim 1-10\,{\rm nHz}$ frequency region. Taking the NANOGrav observational result at face value, we show that 
this signal would be fully consistent with a SGWB produced by an unresolved population of in-spiralling massive black hole binaries (MBHBs) predicted by current theoretical models. Considering an astrophysically agnostic model we find that the MBHB merger rate is loosely constrained to the range $10^{-11} - 2$ $\mathrm{Mpc}^{-3}\,\mathrm{Gyr}^{-1}$. Including additional constraints from galaxy pairing fractions and MBH-bulge scaling relations, we find that the MBHB merger rate is $10^{-5} - 5\times10^{-4}$ $\mathrm{Mpc}^{-3}\,\mathrm{Gyr}^{-1}$, the MBHB merger time-scale is $\le 3\,\mathrm{Gyr}$ and the norm of the $M_\mathrm{BH}-M_\mathrm{bulge}$ relation $\ge 1.2\times 10^{8}\,M_\odot$ (all intervals quoted at 90\% confidence).  
Regardless of the astrophysical details of MBHB assembly, this result would imply that a sufficiently large population of massive black holes pair up, form binaries and merge within a Hubble time.
\end{abstract}



\section{Introduction}

Accurate, decade-long timing of an ensemble of milli-second pulsars -- a pulsar timing array (PTA) -- provides a means to detect gravitational waves in the nanohertz frequency band~\cite{1978SvA....22...36S, 1979ApJ...234.1100D, FosterBacker:1990}. Over the last twenty years, several pulsar timing arrays (PTAs) have been used to achieve ever increasing sensitivity. One of the possible signals that could emerge from such timing campaigns is an isotropic, Gaussian stochastic gravitational-wave background (SGWB). As PTAs' peak sensitivity is sufficiently narrow-band, in this frequency range the SGWB signal can be simply cast in the form
\begin{equation}
    h_c(f) = \Ayr \left(\frac{f}{1\,\mathrm{yr}^{-1}}\right)^\alpha\,,
    \label{eqn:hc0}
\end{equation}
where $h_c$ is the characteristic gravitational wave (GW) strain amplitude, $\Ayr$ is its (unknown) amplitude at the reference frequency of $1\,\mathrm{yr}^{-1}$ and $\alpha$ is the (unknown) spectral slope, which is related to the timing residuals power spectral density, $\gamma$, by the relation $\gamma=3-2\alpha$. 

To date several upper limits on the SGWB have been placed by PTA groups. The most stringent upper limit is from the Parkes PTA~\citep[PPTA,][]{ShannonEtAl:2015}, $\Ayr \leq 1\times 10^{-15}$ with $95\%$ confidence. The European PTA ~\citep[EPTA][]{LentatiEtAl:2015}, the North American Nanohertz Observatory for Gravitational Waves~\citep[NANOGrav][]{ArzoumanianEtAl:2018}, and the International PTA~\citep[IPTA][]{VerbiestEtAl:2016} have also reported comparable upper limits, just a factor of $\approx 2$ higher. However,  NANOGrav have recently reported evidence for a common stochastic red process with amplitude $1.37-2.67\times 10^{-15}$ (5\%--95\% quantiles) 
across 45 pulsars timed in their 12.5 year data set~\citep{ArzoumanianEtAl:2020}, but with insufficient evidence for a Hellings and Downs correlation~\citep{HellingsDowns:1983}, which would be necessary to claim a GW origin of the signal. The odds of a common vs independent signal are $\sim 10^3-10^4:1$, depending on the assumptions in the analyses. This detection is in tension with previously published upper-limits by NANOGrav~\citep{ArzoumanianEtAl:2018} and, more noticeably PPTA \citep{ShannonEtAl:2015}, but~\cite{ArzoumanianEtAl:2020} and \cite{2020arXiv200905143H} suggest that problems with the ephemeris and noise model in previous analyses may have conspired to underestimate the upper-limit value on the amplitude that were previously reported.

Taking the NANOGrav result at face value, several authors have recently argued that a number of processes such as first-order phase transitions, cosmic strings, domain walls, large amplitude curvature perturbations, primordial black holes, inflation can indeed produce a SGWB with amplitude and spectral index consistent with this result.

Since the early days, one of the main justifications for pursuing PTAs as GW detectors is the astrophysical scenario of a SGWB produced by the incoherent super-position of gravitational radiation from adiabatically in-spiraling massive black hole binaries (MBHBs) at the centre of galaxies. 
Individual massive black holes (MBHs) exist at the centres of most galaxies~\citep{KormendyHo:2013}, and hierarchical galaxy formation scenarios indicate that galaxy mergers are frequent throughout cosmic time~\citep{WhiteRees:1978,BegelmanBlanfordRees:1980}. It is therefore natural to conceive a population of MBHBs emitting GWs in the nanohertz frequency band~\citep{1995ApJ...446..543R, Jaffe:2002rt, 2008MNRAS.390..192S, SesanaGWSBHB:2013}. We first consider Monte-Carlo realisations of the SGWB from~\cite{2015MNRAS.451.2417R} to demonstrate the consistency of the expected spectral properties of this signal with the constraints placed by the NANOGrav analysis. We then perform Bayesian inference on parametric models of the SGWB signal based on two sets of assumptions for the underlying MBHB population: an agnostic phenomenological population model in which binaries are assumed in circular orbits and on which we impose minimal prior constraints following ~\cite{MiddletonEtAl:2016}; and an astrophysically driven model which accounts for binary eccentricity as in~\cite{ChenSesanaDelPozzo:2017, ChenEtAl:2017, MiddletonEtAl:2018}, and includes coupling with the environment and additional constraints from independent observations regarding galaxy pair fraction and $M_\mathrm{BH}-M_\mathrm{bulge}$ relation following~\cite{ChenSesanaConselice:2019}.

The paper is organised as follows: in Section~\ref{sec:mcpop} we compare the SGWB expected form Monte Carlo realizations of the MBHB population to the NANOGrav measurement. In Section~\ref{sec:inference} we review the models employed for constructing the SGWB from the MBHB cosmic population and the method for astrophysical inference on the model parameters. Results of our Bayesian inference are presented in Section~\ref{sec:results} and our main findings are summarized in Section~\ref{sec:conclusions}.

\section{Gravitational wave signal from MBHBs}
\label{sec:mcpop}

We start by showing that the results reported by NANOGrav are consistent with existing theoretical predictions. Assuming MBHBs have circular, GW driven orbits, the characteristic strain of the SGWB is computed by integrating over the population distribution in redshift, $z$, and chirp mass, $\Mchirp$, where $\Mchirp$ is a combination of the binary component masses $m_1$, $m_2$ given by $\Mchirp = (m_1 m_2)^{3/5} (m_1+m_2)^{-1/5}$, according to~\citet{Phinney:2001}:
\begin{eqnarray}
    h_{\rm c}^2(f) &=& \frac{4 G^{5/3}}{3 \pi^{1/3} c^2} f^{-4/3} \int {\rm d} \Mchirp \int {\rm d} z \nonumber\\
    & & \times (1+z)^{-1/3} \Mchirp^{5/3} \frac{{\rm d}^2 n}{{\rm d}z {\rm d} \Mchirp}\,,
    \label{eqn:hc}
\end{eqnarray}
where $G$ and $c$ are the gravitational constant and speed of light respectively. 
The population of sources is described by the distribution ${\rm d}^2 n / ({\rm d}z {\rm d} \Mchirp$) as the number density  of binaries per unit redshift, and chirp mass interval.

Eq. \eqref{eqn:hc} expresses the SGWB in terms of its average energy density as a function of frequency, and results by construction in a continuous, smooth spectrum. In reality, the signal is given by an incoherent superposition of discrete quasi-monochromatic sources that can significantly depart from isotropy, Gaussianity and even result in individually resolvable systems \citep{2009MNRAS.394.2255S,2013CQGra..30v4005C,2020PhRvD.102h4039T}, and its frequency spectrum is resolved in bins $\Delta{f_i}=1/T$, where $T$ is the baseline of the observing PTA. A practical way to compute the actual signal spectrum is the following: convert ${\rm d}^2 n/({\rm d}z {\rm d}\mathcal{M})$ into a $
{\rm d}^3 N/({\rm d}z{\rm d} \mathcal{M}{\rm d}f)$ -- i.e., the distribution of sources emitting per unit chirp mass, redshift and frequency -- make a Monte Carlo draw from this distribution, and add up the signal from each individual binary to obtain
\begin{equation}
  h^2_c(f_i)=\frac{\sum_k h^2_kf_k}{\Delta{f_i}}.
  \label{eq:hcmc}
\end{equation}
Here $\Delta{f_i}$ is the $i$-th frequency bin over which the GW spectrum is resolved, the sum runs over all the $k$ sources emitting in the $i$-th frequency bin and $h_k$ is the inclination-polarization averaged strain emitted by each individual MBHB in the Monte Carlo drawing. Full details of this procedure can be found in \cite{2008MNRAS.390..192S}.

\begin{figure}
    \centering
        \includegraphics[width=0.45\textwidth]{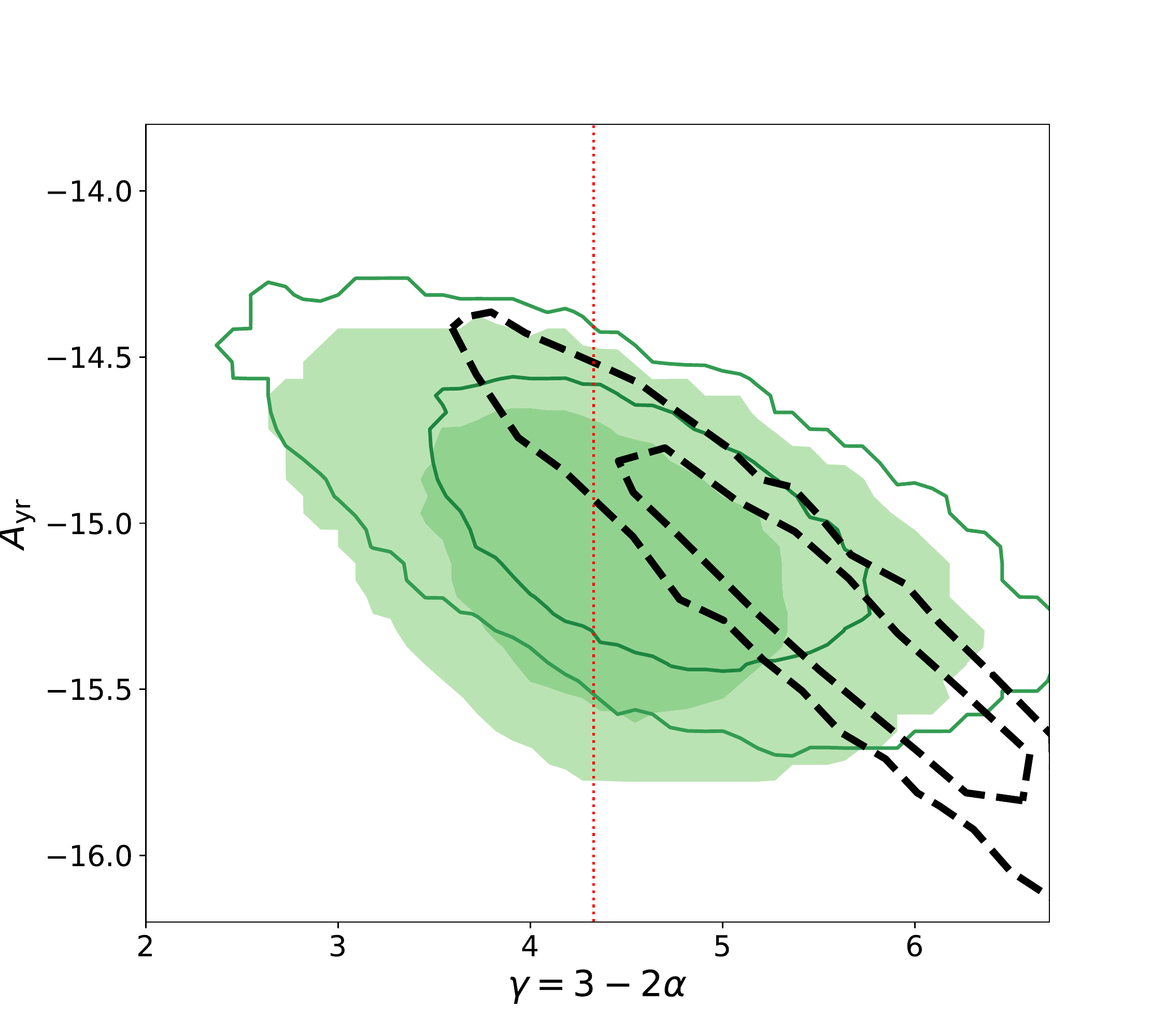}
    \caption{$68\%$ and $95\%$ credible region of the amplitude and slope $(\Ayr,\gamma)$ of the common red signal detected by NANOGrav (black contours) and predicted by the Monte Carlo MBHB population \protect\cite{2015MNRAS.451.2417R} (green contours). Filled contours are for the complete suite of 234K models while open contours are only for models featuring the revised MBH-host relations (see main text for details). The dashed vertical line indicates $\gamma=13/3$.}
    \label{fig:gamma}
\end{figure} 

We consider here the 234K Monte Carlo MBHB populations presented in \cite{2015MNRAS.451.2417R}. For each realisation of the population, we compute the predicted SGWB strain according to Eq. \eqref{eq:hcmc}, using the same binning of the NANOGrav 12.5-yr power spectrum (i.e., $\Delta{f_i}=f_{i+1}-f_i$, where $f_i=i/(12.5 {\rm yr})$). We then make a least square fit to a single power-law for $h_c$ and construct 234K pairs of coefficients $(\Ayr,\gamma)$, which we compare with the NANOGrav results.

An interesting feature of the NANOGrav inference is that the posterior of the inferred spectral index is centred at $\gamma>13/3$. This feature has prompted a number of interpretations of the signal related to an early Universe origin, which would produce steeper red spectra compared to a population of MBHBs. Fig. \ref{fig:gamma} shows probability density contours of $(\Ayr,\gamma)$ obtained from the 234K SGWB realizations of \cite{2015MNRAS.451.2417R}. Although centred around $\gamma=13/3$, the 95\% credible region extends in the range $3\lesssim\gamma\lesssim6$, and there is a significant overlap in the 2D 68\% credible region predicted by the models and measured by NANOGrav. The models were constructed employing a large set of MBH-host relations published in the literature over two decades. In particular, the discovery of overmassive black holes in bright cluster ellipticals, together with the amendment of several dynamical individual MBH mass measurements, prompted an upward revision of the MBH-bulge relations \citep{graham12,mcconnell12,KormendyHo:2013}. If we limit our set to those models, the expected $(\Ayr,\gamma)$ contours shift to the upper right, showing an even higher level of consistency with the published NANOGrav posterior.

\section{SGWB modelling and astrophysical inference method}
\label{sec:inference}

Now that we have established the consistency of the NANOGrav result with theoretical predictions about MBHB populations, we can consider parametric models describing this population and employ Bayesian inference to explore the implications and constraints on the underlying astrophysical model parameters. For this study, we consider two specific population models.

The first is the model developed in \cite{MiddletonEtAl:2016}, and we refer to it as M16 hereafter. Binaries are assumed to be circular and GW driven and the strain of the SGWB is fully described by Eq.~\eqref{eqn:hc}, where we make only minimal assumptions about the underlying MBHB population. In fact, the distribution ${\rm d}^2 n / ({\rm d}z {\rm d} \Mchirp)$, i.e., the number density  of binaries per unit redshift, and chirp mass interval, is modelled as a Schechter function of the form \citep{1976ApJ...203..297S}:
\begin{eqnarray}
    \frac{{\rm d}^2 n}{{\rm d}z {\rm d} \log_{10} \Mchirp} =& \nonumber 
    \no \left[ \left( \frac{\Mchirp}{10^7\Msol}\right)^{-\alphaM} \exp^{-\Mchirp/\Mstar} \right] \\
    & \times \left[ (1+z)^{\betaz} \exp^{-z/z_0} \right] 
    \frac{{\rm d}t_{\rm R}}{{\rm d}z}\,,
    \label{eqn:model}
\end{eqnarray}
where $t_{\rm R}$ is the time in the rest frame of the source and we assume values for the cosmological parameters according to  $H_0 = 70\,{\rm km}\,{\rm s}^{-1}\,{\rm Mpc}^{-1}$ (i.e. $h_0=0.7$), $\Omega_{\rm M} = 0.3$, $\Omega_{\Lambda}=0.7$, and $\Omega_{\rm k}=0$.

The shape and magnitude of the population distribution is described by five parameters: $\theta = \{\no,\alphaM,\Mstar,\betaz,z_0\}$, where $\{\alphaM,\Mstar\}$ and $\{\betaz,z_0\}$ control the shape of the $\Mchirp$ and $z$ distributions respectively, and $\no$ is the merger rate per unit rest-frame time, comoving volume and logarithmic chirp mass interval.
The integration limits in Eq.~\ref{eqn:hc} are set to $0\leq z \leq 5$ and $10^6 \leq \Mchirp/\Msol \leq 10^{11}$. 

We then consider a second model following~\cite{ChenSesanaConselice:2019}, to which we will refer as C19 hereinafter. In this model the MBHB population is allowed to be eccentric and to interact with the environment, which can cause a departure of the signal from the single power-law described by Eq.~\eqref{eqn:hc}. In this model, the characteristic amplitude $h_\mathrm{c}(f)$ is described by 18 independent parameters related to the galaxy stellar mass function (described by a redshift evolving single Schechter function), the black hole pair fraction, merger timescale and galaxy -- MBH scaling relation. We refer the reader to ~\cite{ChenSesanaConselice:2019} for full details. In summary, the MBHB merger rate $d^2 n/(dzd \mathcal{M})$ is described by 16 parameters related to astrophysical observables: $(\Phi_0, \Phi_I, M_0, \alpha_0, \alpha_I)$ for the galaxy stellar mass function, $(f_0', \alpha_f, \beta_f, \gamma_f)$ for the pair fraction, $(\tau_0, \alpha_\tau, \beta_\tau, \gamma_\tau)$ for the merger timescale, and $(M_*,\alpha_*, \epsilon)$ for the $M_\mathrm{BH}-M_\mathrm{bulge}$ scaling relation. The first three sets of parameters can be combined into five effective parameters that allow to write the merger rate in the form given by Eq.~\eqref{eqn:model}, to ease comparison with the agnostic M16 model. Finally, two extra parameters, $\zeta_0$ and $e_0$, describe the density of the stellar environment and the eccentricity at the MBHB pairing respectively, and can produce a low frequency departure of the spectrum from the single power-law of Eq.~\eqref{eqn:hc}. 

We use Bayesian inference to find the posterior distribution $p(\theta|d,M)$ for the parameters $\theta$ of model $M$ given the observation data $d$.
\begin{equation}
 p(\theta|d,M) = \frac{p(\theta| M) p(d|\theta,M)}{p(d|M)}\,,
 \label{eqn:bayes}
\end{equation}
where $p(\theta| M)$ is the prior distribution on the model parameters, $p(d|\theta,M)$ is the likelihood of model $M$ with parameters $\theta$ of producing the data, and $p(d|M)$ is the evidence.

We assume the detected strain spectrum to be log normally distributed at different frequency bins $f$, thus the likelihood takes the general form
\begin{equation}
\log_{10} p\left(d|\theta,M \right) \propto - \frac{1}{2}\sum_f \frac{\left[ \log_{10}A(f; {\theta,\,M})-\log_{10}d(f)\right]^2}{\sigma^2(f)} \,,
\label{eq:likedet}
\end{equation}
where each frequency bin can be treated independently with its own detected strain amplitude $d(f)$ and uncertainty $\sigma(f)$. The predicted amplitude $A(f; {\theta,\,M})$ from a given model $M$ and parameters $\theta$ is then compared against the detected amplitudes from the data $d(f)$.

In our inference scheme, the data $d(f)$ would ideally be the posterior chains of the NANOGrav analysis of the $12.5\,{\rm yr}$ data set~\citep{ArzoumanianEtAl:2020}. Since those are not publicly available, however, we construct analytical proxies for the posterior distribution of the detected signal. The M16 model take as single data input the measured signal amplitude at $f=1\,$yr$^{-1}$. We model it as a log normal distribution centred around ${\rm log}_{10}\Ayr=-14.7$ with $\sigma=0.09$, which reproduces the  $5$--$95\%$ credible range in $\Ayr$ of $1.37$--$2.67 \times 10^{-15}$ quoted by NANOGrav. In order to use consistent input data for the C19 model, we simulate a detection of the same nominal amplitude and uncertainty assuming a population of MBHBs in circular orbits ($\alpha=-2/3$ power-law) and ${\rm log}_{10}\Ayr=-14.7$. We match the frequency range of the five lowest frequency bins $f_i=i/T$ used in the `free spectrum' analysis reported by NANOGrav, compute the strain value and assign a Gaussian uncertainty of $\sigma_i=0.09$ at each frequency bin.

Next, we need to specify the priors. For the M16 model the priors are identical to those used in~\cite{MiddletonEtAl:2016} with the exception of $\Mstar$ which is now uniform in the range $\log_{10}\Mstar\in[6,10]$ and $\no$ which is now uniform in the range $\logno\in[-20,3]$. 
For the C19 model we use the `extended' prior range reported in Table 1 of  ~\cite{ChenSesanaConselice:2019}. We note that with this choice of priors, the predicted range of the SGWB amplitude of the C19 model is consistent with the Monte Carlo models of  \cite{2015MNRAS.451.2417R} considered in Section ~\ref{sec:mcpop}.

Finally, we compute posterior density distributions on the model parameters using \texttt{cpnest}, a nested sampler ~\citep{cpnest, VeitchVecchio:2010} and \texttt{PTMCMCSampler}, a Markov Chain Monte Carlo based sampler \citep{ptmcmc}.

\section{Results: implications for the MBHB population}
\label{sec:results}

We discuss here the most relevant features of the M16 and C19 analysis; full corner plots showing the posterior distributions of all model parameters are shown in Appendix \ref{app:cornerPlot}.

\begin{figure}
    \centering
    \includegraphics[width=0.48\textwidth]{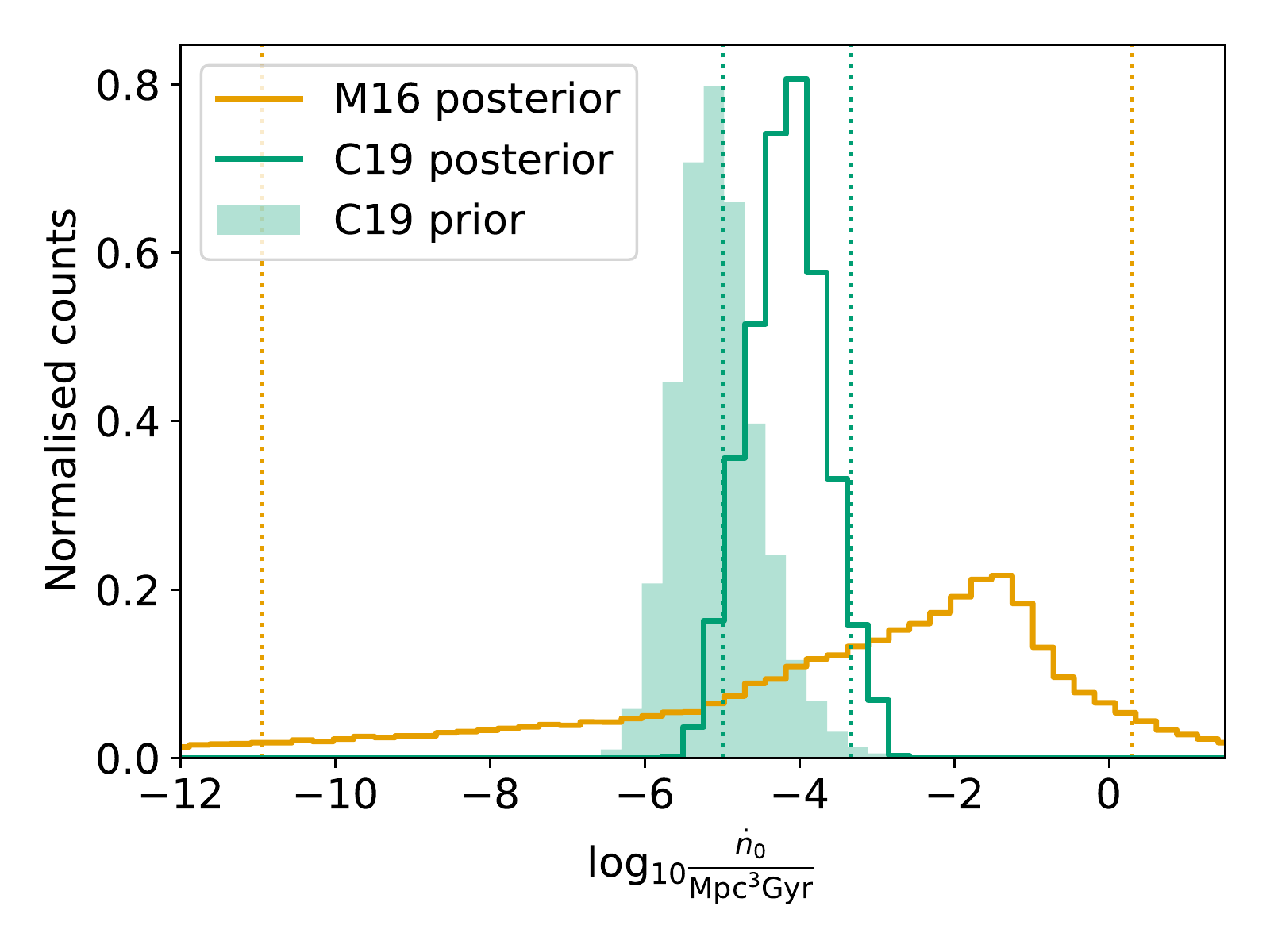}
    \caption{Marginalised posterior distribution for $\no$. 
            The orange and green outlines show the posterior distributions for the M16 and C19 models respectively. 
            The filled green histogram shows the prior for C19 (the prior for M16 is not shown as it is uniform in the range $[-20,3]$).
            The vertical dotted lines indicate the $5\%$ and $95\%$ percentiles. }
  \label{fig:logno}
\end{figure}

In the agnostic M16 models, the SGWB measurement mostly constrains the MBHB number density, described by the parameter $\logno$. All other parameters return essentially flat one-dimensional priors (see Fig. \ref{fig:corner}). Constraining only a single parameter is unsurprising given we are using a single observational constraint in our likelihood computation. The posterior distribution for $\logno$ is shown in Fig.~\ref{fig:logno}. We find a central $90\%$ credible region ($5$--$95\%$) of $\lognMLowPost$ to $\lognMHighPost$ in $\logno$ 
This is compared to the posterior on the equivalent effective parameter in the C19 model. The latter is constrained to a much narrower range by the astrophysical prior, and tends to push to its higher bound. The distribution peak is at $\logno\approx -4$, offset from the M16 model. Still, there is full consistency between the two as the C19 posterior support is fully included within the M16 one.

\begin{figure}
    \centering
    \includegraphics[width=0.48\textwidth]{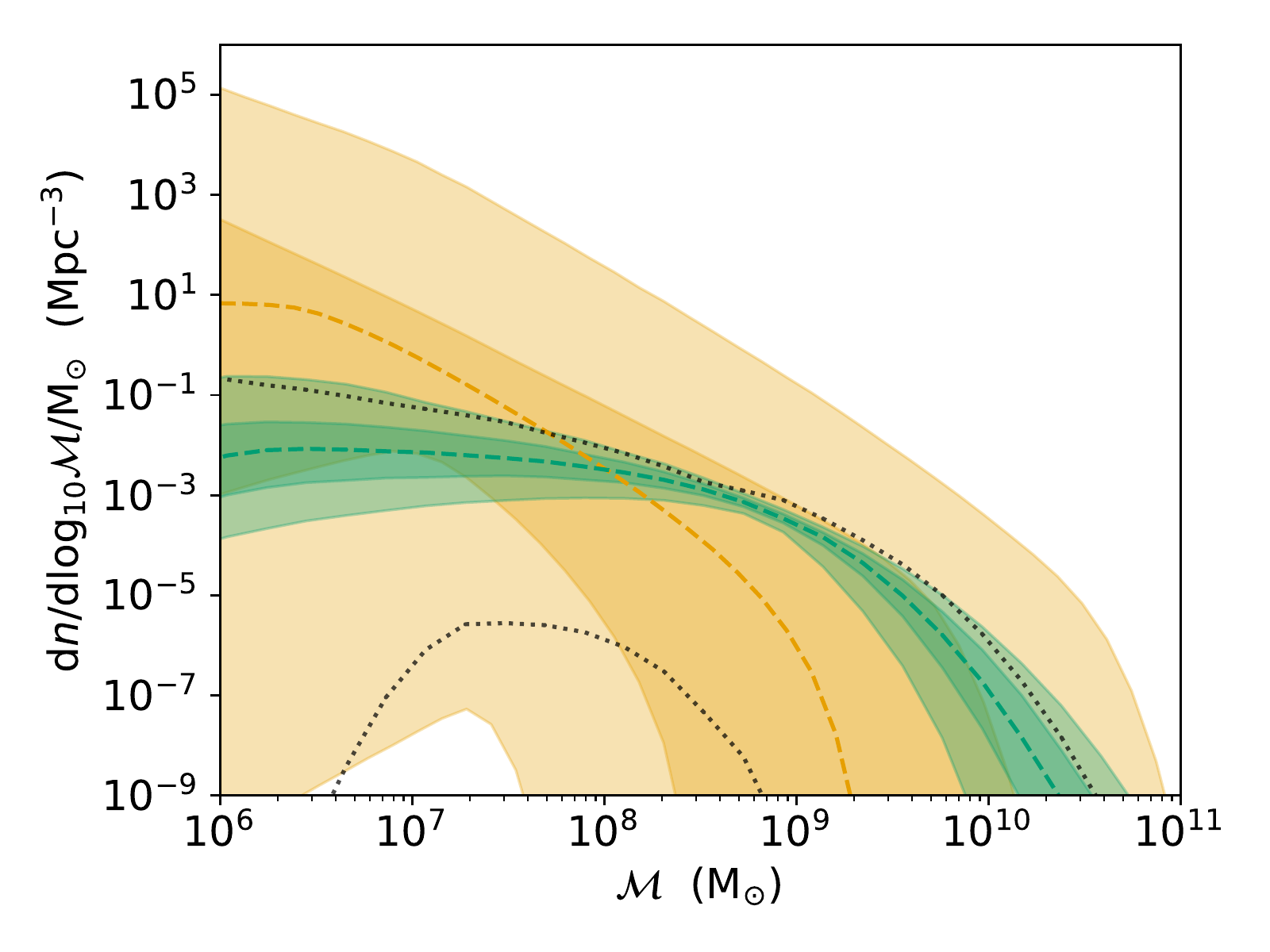}
    \caption{Merger rate density vs chirp mass posterior for the M16 and C19 analysis in orange and green respectively. The shaded regions show the central $50\%$ and $90\%$ credible regions and the dashed lines show the median. 
    The black-dotted lines show the central $99\%$ region for the C19 prior. 
    }
    \label{fig:conf_m}
\end{figure}

The difference in the $\logno$ is explained by the inference on the redshift-integrated MBHB mass function, shown in Fig. \ref{fig:conf_m}. The astrophysical prior used in the C19 model leaves little room for a steep negative slope of the MBHB chirp mass function below ${\cal M}\approx 10^8\,{\rm M}_{\odot}$. However, no such restriction is imposed in the M16 model. The low mass end of the MBHB mass function can be quite steep, resulting in a much higher integrated merger rate (i.e. a higher $\logno$). 

\begin{figure}
    \centering
    \includegraphics[width=0.48\textwidth]{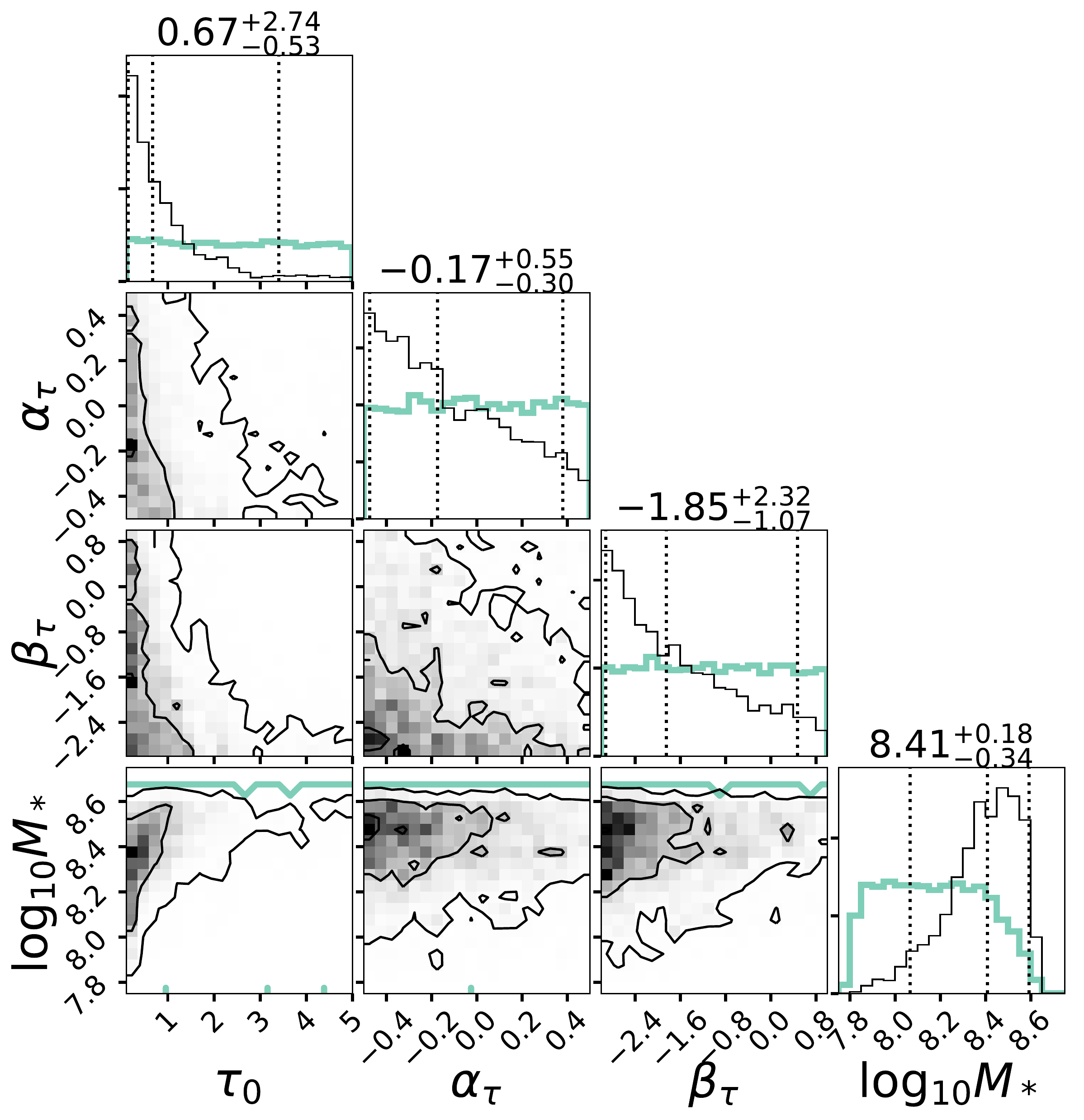}
    \caption{Corner plot showing 2D posterior distributions for selected parameters of the C19 models. In the panels showing 1D marginalized distributions,  posteriors are shown as black histograms, while priors are shown as green histograms.}
    \label{fig:histv}
\end{figure}

Finally, we can look at whether the NANOGrav measurement constrains any other interesting astrophysical parameter of the C19. Contrary to the M16 model, the C19 one does not start from a coalescing MBHB function, but from an astrophysically constrained population of galaxy pairs. Within these galaxies reside MBHs that might eventually merge following the galaxy merger. The formation and coalescence of the MBHBs is not postulated, but it is bound to a coalescence timescale, which is described by a function of the form
\begin{equation}
\tau(M,z,q) = \tau_0 \Big( \frac{M}{b M_0} \Big)^{\alpha_\tau} (1+z)^{\beta_\tau} q^{\gamma_\tau}
\label{tau}\,,
\end{equation}
where $\tau_0$ is an (unknown) constant which sets the overall time-scale, $M$ is the mass of the primary galaxy, $q<1$ is the galaxy mass ratio and $b M_0 = 0.4/h_0 \times 10^{11} \Msol$. 
$\tau$ describes the typical time elapsed between the two galaxies being at a (projected) distance of $\approx 30$kpc, where galaxy pairs are counted, and the final coalescence of the MBHB, including pairing via dynamical friction, hardening and final GW emission. Constraints on $\tau_0, \alpha_\tau, \beta_\tau$ are shown in Fig. \ref{fig:histv}. The data favour rapid coalescence of the binary pairs, with characteristic $\tau_0<3$ Gyr at 90\% confidence. The fact that both $\alpha_\tau, \beta_\tau$ are skewed toward negative values suggests that short merger timescales for massive black holes are preferred. In the C19 models, the black hole mass $M_{\rm BH}$ is connected to the galaxy bulge mass $M_{\rm bulge}$ via a scaling relation of the form \citep[see, e.g., ][]{KormendyHo:2013}

\begin{equation}
M_{\rm BH} = {\cal N}\left\{M_* \Big( \frac{M_{\rm bulge}}{10^{11} M_\odot} \Big)^{\alpha_*}, \epsilon\right\},
\label{eqnmbulge}
\end{equation}
where ${\cal N}\{x,y\}$ is a log normal distribution with mean value $x$ and standard deviation $y$. The measurement of a SGWB with $\Ayr\approx 2\times 10^{-15}$ naturally favours high normalizations in Eq. \eqref{eqnmbulge}, resulting in a clear preference for high $M_*$, as shown in the bottom-right panel of Fig. \ref{fig:histv}.

\section{Conclusions}
\label{sec:conclusions}

We performed a systematic analysis of the astrophysical implications of the reported NANOGrav common stochastic red process~\cite{ArzoumanianEtAl:2020}, under the assumption that this signal is of astrophysical origin and due to a SGWB from a population of MBHBs. Our analysis consists of several steps, resulting in a consistent picture.

First, we use forward modelling of the SGWB signal from the suite of Monte Carlo realization of the MBHB population presented in \cite{2015MNRAS.451.2417R}, to show that the 2D credible region of the signal amplitude and spectral slope $(\Ayr,\gamma)$ of the common red process reported by NANOGrav is fully consistent with the incoherent superposition of radiation from MBHBs that are individually unresolvable. 
This forward modelling also shows that $M_{\rm BH}-M_{\rm bulge}$ relations with high normalizations better reproduce the NANOGrav signal.

Then, by means of Bayesian inference on a parametric agnostic model of the MBHB population \citep{MiddletonEtAl:2016}, we show that, without any other prior information, the detected signal implies a MBHB merger rate density in the range $10^{-11} - 2$ $\mathrm{Mpc}^{-3}\,\mathrm{Gyr}^{-1}$ (90\% confidence), which is consistent with expectations from independent estimates of galaxy and MBHB merger models.

Finally, we use the observation based parametric model of \cite{ChenSesanaConselice:2019}, which makes use of prior knowledge on the galaxy mass function and pair fraction, expected merger timescale and the observed $M_{\rm BH}-M_{\rm bulge}$ relation, to explore whether the signal places any further constraint on any of those observables. We find that the SGWB measurement does not add information to the majority of the parameters defining the model (see Appendix \ref{app:cornerPlot}), with the notable exceptions of $\tau_0, \alpha_\tau, \beta_\tau$, defining the MBHB merger timescale and $M_*$, defining the normalization of the MBH-host bulge mass relation. The measurement of a SGWB with $\Ayr\approx 2\times 10^{-15}$ naturally favours high normalisations in the $M_{\rm BH}-M_{\rm bulge}$ relation, consistent with recent estimates in the literature \citep[e.g.][]{mcconnell12,KormendyHo:2013}, and relatively short merger timescales, as implied by the preference for short $\tau_0$ and negative $\alpha_\tau, \beta_\tau$. 

These results demonstrate that the common stochastic red process reported by NANOGrav is fully consistent with our current knowledge of the cosmic MBHB population, although the measurement is also consistent with a steeper red spectrum centred around $h_c\propto f^{-1}-f^{-1.5}$. In particular, this latter feature has prompted a number of early Universe interpretations of the signal, including first order phase transitions, cosmic strings, domain walls, large amplitude curvature perturbations, primordial black holes, inflation, etc. In light of our findings however, if future PTA analyses on longer and more sensitive data sets, including a larger number of pulsars provided stronger statistical significance for the detection of this signal, the MBHB scenario would offer a natural explanation without the need of invoking more exotic physical processes.  Most importantly, the signal would provide direct evidence that MBHB mergers occur in Nature.

As pulsar timing campaigns from PTA groups around the world are continuously providing more and higher sensitivity data, it will be interesting to see how the statistical significance of a common stochastic signal across a large array of pulsars evolves and, crucially, if it reveals a Hellings and Downs correlation expected from GWs. If, in fact, we are seeing the first glimpse of a GW signal, the next few years will be decisive in assessing its nature \citep{2020arXiv201011950P}.

\section*{Acknowledgements}
Parts of work were performed on the OzSTAR national facility at Swinburne University of Technology. The OzSTAR program receives funding in part from the Astronomy National Collaborative Research Infrastructure Strategy (NCRIS) allocation provided by the Australian Government. H.M. is supported by the Australian Research Council Centre of Excellence for Gravitational Wave Discovery (OzGrav) (project number CE170100004). A.S. is supported by the European Research Council (ERC) under the European Union’s Horizon 2020 research and innovation program ERC-2018-COG under grant agreement No 818691 (B Massive). S.C. acknowledges support from the CNRS, CEA, CNES in France. A.V. acknowledges support from the Royal Society and the Wolfson Foundation. The Authors acknowledge support from EPTA, NANOGrav, PPTA and IPTA.

\section*{Data Availability}
The posteriors samples used to produce this work are available from this link \url{https://github.com/hannahm8/PTAInference}.



\bibliographystyle{mnras}
\bibliography{mbhb} 




\appendix

\section{Corner plots}
\label{app:cornerPlot}
\begin{figure*}
    \centering
    \includegraphics[width=0.9\textwidth]{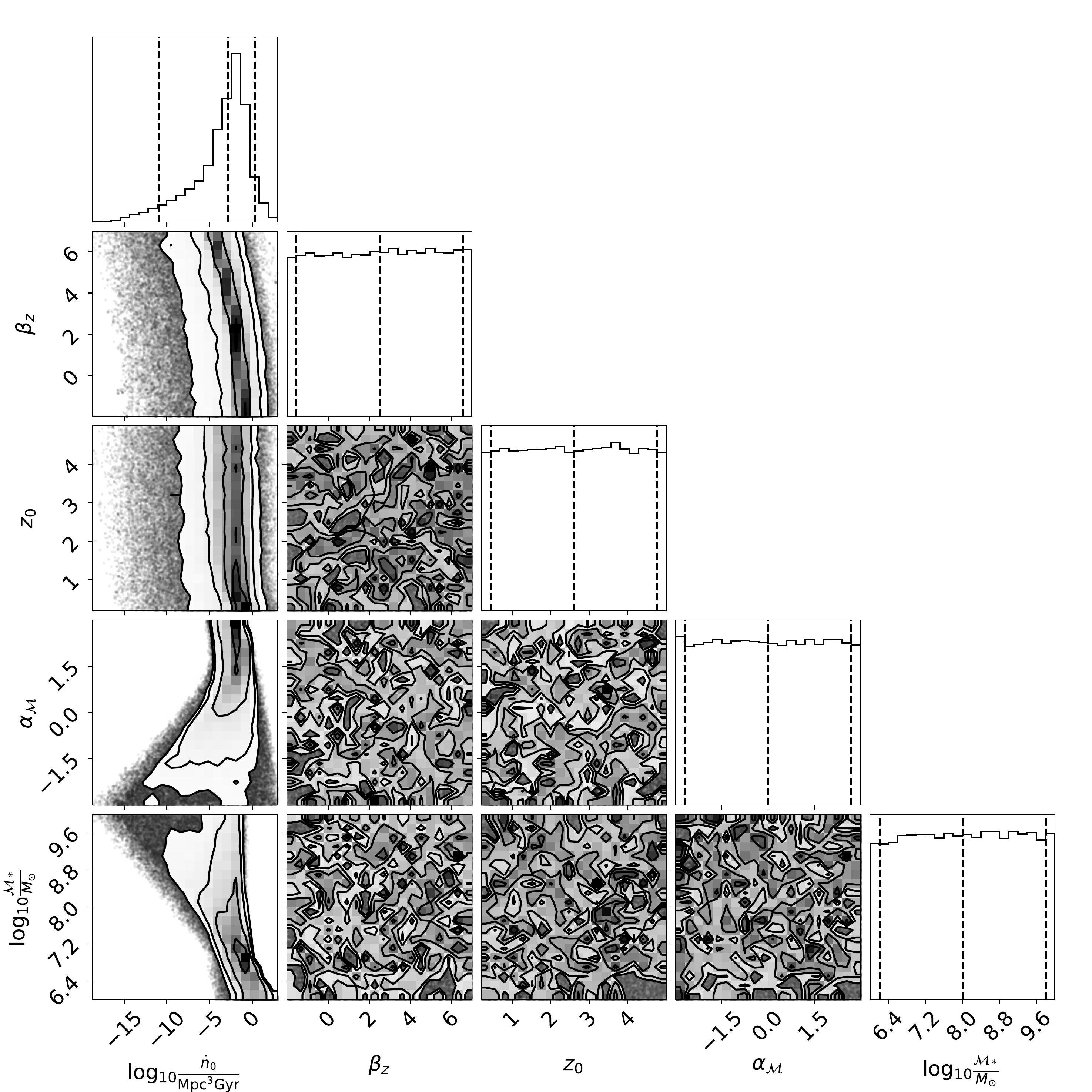}
    \caption{Marginalised posterior distributions for the five parameter M16 model.
    The central plots show the two dimensional posterior for the parameter combinations and the histograms show the one dimensional posterior for each parameter. 
    The vertical dotted lines indicate the $5\%$, $50\%$, and $95\%$ percentiles.}
    \label{fig:corner}
\end{figure*}


\begin{figure*}
    \centering
    \includegraphics[width=0.9\textwidth]{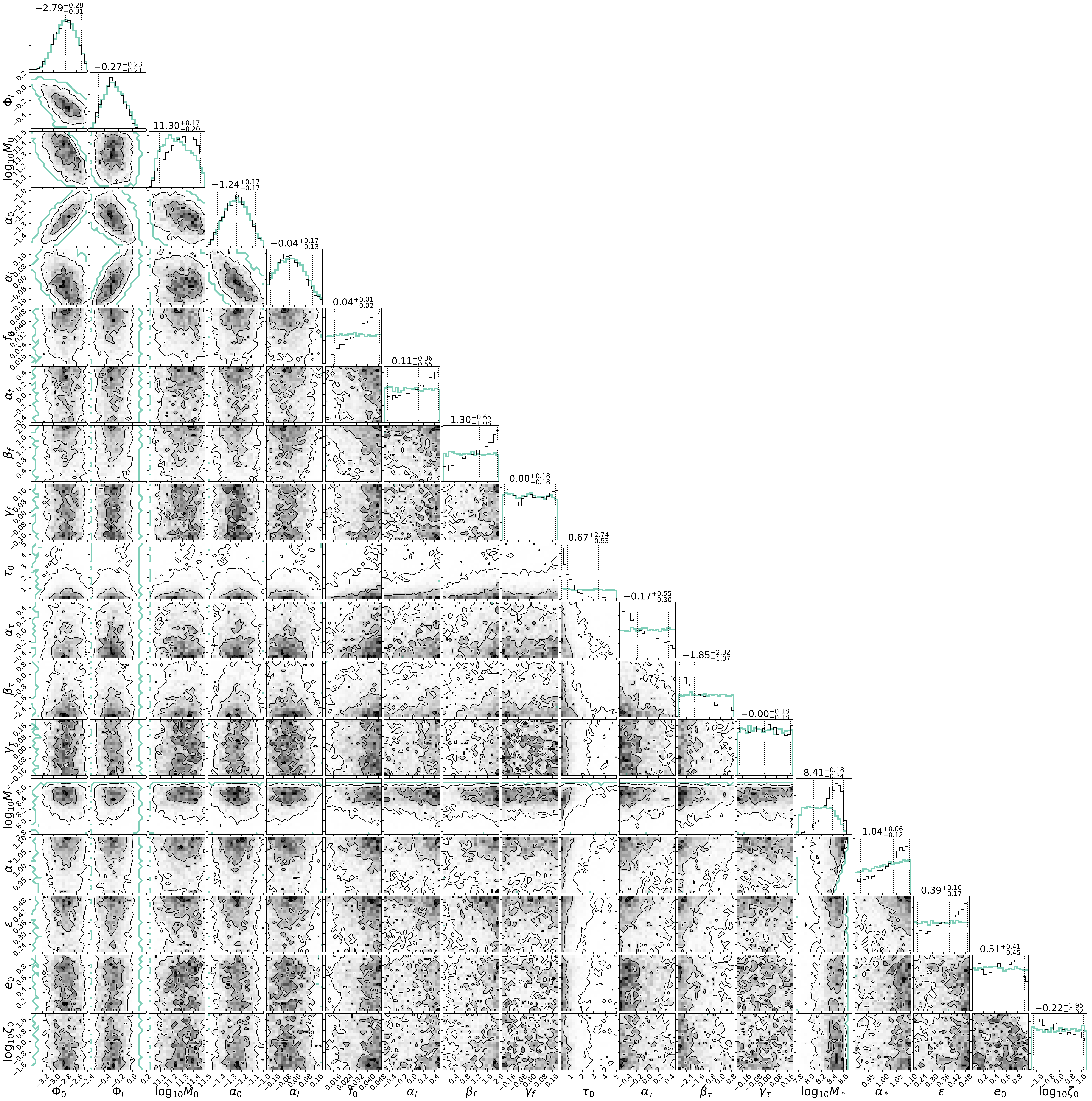}
    \caption{Same as Fig. \ref{fig:corner} but for the 18 parameters defining the C19 model.}
    \label{fig:corner_gal_ext}
\end{figure*}


\bsp	
\label{lastpage}
\end{document}